\documentclass[runningheads,a4paper]{llncs}

\usepackage{algorithmic}
\usepackage{graphicx}
\usepackage{textcomp}
\usepackage{xcolor}
\usepackage{xurl}
\PassOptionsToPackage{hyphens}{url}\usepackage{hyperref}
\usepackage{float}
\usepackage{multirow}
\usepackage{multicol}
\usepackage[frozencache,cachedir=.]{minted}

\usepackage{tabularx}

\begin{document}

\author{Pierrick Pochelu}
\institute{University of Luxembourg, FSTM-DCS \\
\email{pierrick.pochelu@uni.lu}}

\authorrunning{P. Pochelu}

\title{JaxDecompiler: Redefining Gradient-Informed Software Design}
\maketitle

\begin{abstract}
Among numerical libraries capable of computing gradient descent optimization, JAX stands out by offering more features, accelerated by an intermediate representation known as Jaxpr language. However, editing the Jaxpr code is not directly possible. This article introduces JaxDecompiler, a tool that transforms any JAX function into an editable Python code, especially useful for editing the JAX function generated by the gradient function. JaxDecompiler simplifies the processes of reverse engineering, understanding, customizing, and interoperability of software developed by JAX. We highlight its capabilities, emphasize its practical applications especially in deep learning and more generally gradient-informed software, and demonstrate that the decompiled code speed performance is similar to the original. 
\end{abstract}

%\begin{IEEEkeywords}
%Deep learning, neural network, inference system, software optimization
%\end{IEEEkeywords}

% !TeX TXS-program:compile = txs:///pdflatex/[--shell-escape]

\section{Introduction}

% Jax
Computational science and data science are witnessing a profound transformation, marked by the emergence of powerful new numerical frameworks. Recently, numerical frameworks such as Jax \cite{jax:2021} \cite{jax:2018}, Tensorflow \cite{tf:2020}, PyTorch \cite{pytorch:2011} and Sympy \cite{sympy:2017} have transformed the design of mathematical optimization by leveraging symbolic differentiation, eliminating the reliance on numerical approximations. %However, mathematical frameworks do not integrate native language structures like loops, conditional instructions, array indexing, and Numpy API \cite{numpy:2011}.

% Need of JaxDecompiler
% JAX distinguishes itself from PyTorch and Tensorflow by offering a broader range of functionalities. While PyTorch and Tensorflow primarily focus on deep neural network training, JAX not only differentiates functions but also orchestrates their design using native Python language, including loops, indexing, and conditions, Numpy API \cite{numpy:2011} and distributed capabilities based on MapReduce \cite{mapreduce:2008}. In addition to that, advanced derivative capability such as automatic computing of multiple order derivatives based on Autograd \cite{autograd:2015}, which is a lacking functionality of other numerical frameworks primarily designed for deep learning. Finally, JAX relies on low-level Jaxpr code for faster execution, which can be again accelerated with Just-In-Time compiled with  XLA \cite{xla}.  However, the automatically generated Jaxpr code may not always be suitable, necessitating the availability of a decompiler for translating it into Python code for modification, before regenerating Jaxpr for final execution.

JAX distinguishes itself from PyTorch and TensorFlow by offering a broader range of functionalities. While PyTorch and TensorFlow primarily focus on deep neural network training, JAX not only integrates natively Autograd allowing which allows to address optimization problems with multi-order derivatives. The design of the function allows to use of the native Python language, including support for loops, indexing, and conditions. It provides an intuitive Numpy API \cite{numpy:2011} and includes distributed computing capabilities based on the MapReduce programming model \cite{mapreduce:2008}.

%In addition to these features, JAX boasts an advanced derivative capability, enabling automatic computation of multiple-order derivatives based on Autograd \cite{autograd:2015}. This functionality is lacking in other numerical frameworks primarily designed for deep learning. Furthermore, JAX relies on a low-level Jaxpr code for faster execution, which can be further accelerated with Just-In-Time compilation using XLA \cite{xla}. However, it's important to note that the automatically generated Jaxpr code may not always be suitable, highlighting the need for a decompiler to translate it into Python code for modification before regenerating Jaxpr for final execution.

The strength of JAX extends beyond its expressiveness; it also leverages a low-level Jaxpr code for faster execution. This efficiency can be further enhanced through Just-In-Time compilation using XLA \cite{xla}. However, it's crucial to acknowledge that the automatically generated Jaxpr code may not always be suitable for every scenario. This underscores the importance of a decompiler to translate the Jaxpr code into Python, facilitating modification before regenerating Jaxpr for final execution.

% Litterature
Many decompilers have been extensively proposed but they have mostly focused on languages like C, C++, and Java \cite{java:2020}. Those works show that the challenge of designing a decompiler is intrinsically linked to both the source and target language characteristics.

% Benefits of JaxDecompiler
To bridge the gap between Jaxpr and Python and answer this technological gap, we introduce JaxDecompiler. It takes any JAX function as input and produces the equivalent Python code. The decompiler is required when the input function has been generated by a symbolic derivative in the Jaxpr language.

JaxDecompiler may serve a variety of purposes. Decompilers are generally used in applications such as malware detection \cite{malware:2021}, identifying duplicate code \cite{plag:2016}, and offering automatic code design recommendations \cite{log:2021}. JaxDecompilers is also useful for gradient code customization aiming to improve computing speed, and arithmetic stability, or export the gradient code for interoperability. PyTorch, Tensorflow, and JAX \footnote{  \url{https://jax.readthedocs.io/en/latest/notebooks/Custom_derivative_rules_for_Python_code.html}} allows already to customize gradient code by replacing the gradient code of a given function. JaxDecompiler enables another approach to do this, after the gradient code is computed with chain rule and exported as Python with the JaxDecompiler, the user may edit it and have full control of it.

% Usefull for interoperability
%Finally, JaxDecompiler allows to export of the gradient code for interoperability with other software and platforms. PyTorch and Tensorflow rely mostly on the neural network format ONNX, which is less general for extension. It is constrained by the need for a neural network interpreter on the target platform, and primarily target neural network applications.

Finally, JaxDecompiler provides the capability to export Python gradient code, facilitating interoperability with diverse software and platforms. In contrast, PyTorch and Tensorflow use neural network representation syntax for storing them such as ONNX \cite{onnx:} and TorchScript. This is constrained by the requirement for a dedicated neural network interpreter on the target platform and those representations are primarily tailored for neural networks. It's worth noting that, before the introduction of JaxDecompiler, JAX users typically converted their models into Tensorflow, and then from Tensorflow into ONNX for interoperability.

This paper is organized into four main sections. Section~\ref{step} provides practical examples of JaxDecompiler's usage. The inner working of the decompiler is presented in section~\ref{design}. The speed performance of decompiled code is compared to the original JAX code on 3 applications in section~\ref{perf}. Finally, the conclusion in section~\ref{conclusion} summarizes the significance and potential of JaxDecompiler and provides the GitHub link.

\section{Step-by-step use cases}
\label{step}

%In this section, the document provides a detailed explanation of how JaxDecompiler is used, illustrated with a concrete example.

This section provides an example of the typical workflow usage of JaxDecompiler. 

\subsection{Step 1: JaxDecompiler input}

Let's consider a typical JAX code below. In this example, we start with the function \texttt{jnp.log(1+jnp.exp(x))} and aim to obtain the derivative with respect to \texttt{x} encoded as float 32bits. The generated gradient function suffers from arithmetic instability for large \texttt{x} values greater than 87. JaxDecompiler addresses this issue by providing a decompiled version of the JAX function, allowing users to modify the code.

\begin{minted}{python}
from jax import numpy as jnp
def f(x):
    return jnp.log(1+jnp.exp(x))

from jax import grad
gf=grad(f)
print(gf(100.)) # output: nan, expected: 1
\end{minted}

%In this code, a function \texttt{f(x)} is defined, and the derivative of this function is computed using Jax. However, running this code fails when dealing with too large integers.

%TODO gives the link
%The \texttt{grad(f)} function is implemented with JAX and generates a Jaxpr code, which is a low-level representation of the computation. The Jaxpr code is provided using the function \texttt{make\_jaxpr} into JAX \footnote{\url{ https://jax.readthedocs.io/en/latest/_autosummary/jax.make_jaxpr.html }}

The Jaxpr code of \texttt{gf} may be exported:
\begin{verbatim}
{ lambda ; a:f32[]. let
b:f32[] = exp a
c:f32[] = add 1.0 b
_:f32[] = log c
d:f32[] = div 1.0 c
e:f32[] = mul d b
in (e,) }
\end{verbatim}

\subsection{Step 2: Utilizing JaxDecompiler API}

To address the limitations of the Jaxpr code and enable users to better understand, modify, and work with it, JaxDecompiler is introduced. The \texttt{Jaxpr2python} function is a key feature of JaxDecompiler, taking a JAX function as input and returning the decompiled function. The decompiled Python code is also provided as a string for exposing it to the user.

\begin{minted}{python}
from JaxDecompiler import decompiler
gf2,py=decompiler.Jaxpr2python(gf,0.,is_python=True)
\end{minted}

%TODO improve this
%The callable function \texttt{gf2} is equivalent to \texttt{gf} in terms of behaviour, but the implementation of \texttt{gf2 have been decompiled.
The main feature of \texttt{Jaxpr2python} function is to take a function as input (here \texttt{gf2}) and return the decompiled function (here \texttt{gf}) which behaves the same as the input one.  The second argument (here \texttt{0.}) is a fake input used to specify the input type.

The \texttt{is\_python} argument indicates that we return a second output, the decompiled function as a string (here \texttt{py}). Saving and importing \texttt{py} is identical to \texttt{gf}.

The decompiled Python code for the given example is as follows:
\begin{verbatim}
from jax.numpy import *
def gf2(a):
    b = exp(a)
    c = 1.0 + b
    _ = log(c)
    d = 1.0 / c
    e = d * b
    return e
\end{verbatim}

It's important to emphasize that while the equivalence between Jaxpr and Python code may appear straightforward in this example, the JaxDecompiler may handle more intricate patterns, including conditional structures, distributed map operations, and loops. The GitHub repository link at the end provides the opportunity to explore and translate more complex Jaxpr code into comprehensive unit tests and applications, showcasing the decompiler's versatility and utility in handling a wide range of scenarios.

\subsection{Step 3: Consuming JaxDecompiler output}

The generated Python code is owned by the user, providing flexibility to use external code tools or manually edit the code. For example, the code can be manually improved for arithmetic stability with the if block statement to return an approximation:

\begin{verbatim}
from jax.numpy import *
def gf2(a):
    if a>87:
        return 1.
    b = exp(a)
    c = 1.0 + b
    _ = log(c)
    d = 1.0 / c
    e = d * b
    return e
\end{verbatim}

This assembly-style language produced by decompilers is a well-known limitation for human maintenance of large software.
Recent advancements in Large Language Models for processing decompiled codes \cite{dirty:2022} provide optimism for editing decompiler output into equivalent and more human-friendly code.

While the assembly-style Python produced by JaxDecompiler is a limitation when generating large software intended for human maintenance, it is advantageous for transpilation from Python to another language increasing interoperability. The transpiled code from Python into C is given below.

\begin{minted}{cpp}
#include <cmath>
float gf2(float a) {
     if (a > 87) {
         return 1.0;
     }
     float b = exp(a);
     float c = 1.0 + b;
     float d = 1.0 / c;
     float e = d * b;
     return e;
}
\end{minted}

\section{JaxDecompiler design}
\label{design}

This section presents the JaxDecompiler. First, an overview is given, and then the 3 main components are presented in more detail: Tokenizer, Line Translator, and Import Set.

\subsection{Overview} 

\begin{figure}[h]
\centering
  \includegraphics[width=\linewidth , trim=2.8cm 8cm 0 0, clip]{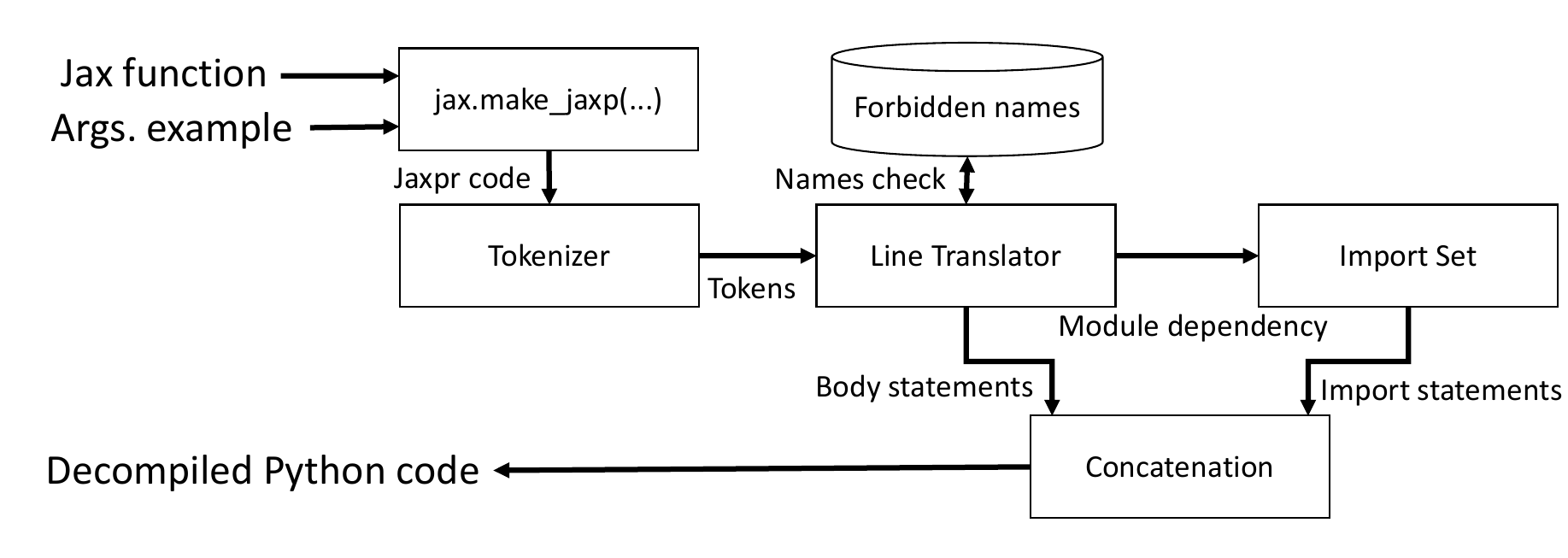} % Replace with your actual filename
  %\vspace{-3.5cm} % Adjust the vertical space here
  \caption{Design of the JaxDecompiler. Edge represents data flow and the box the processing.}
  \label{fig:bigp} % Optional: Add a label for cross-referencing
\end{figure}

The user gives the Jax function, and argument example (to automatically infer data types) and gets the decompiled Python code as output. The flow of data and processes for achieving this is depicted in Figure~\ref{fig:bigp}.

The first step is extracting the JaxPr code from the Jax function using \texttt{make\_jaxpr} \footnote{\url{ https://jax.readthedocs.io/en/latest/_autosummary/jax.make_jaxpr.html }}. The Tokenizer splits Jaxpr lines into tokens, the Line Translator produces body statements, and the Import Set generates Python's import statements.

% Parle
\subsection{Tokenizer} 

Unlike compilers, decompilers take well-formatted code as input, making the input code trivial to analyze. The lexer step splits each Jaxpr statement into 4 sub-parts: 
\begin{enumerate}
    \item From 0 to $n$ output variable name(s).
    \item The operator.
    \item The operator arguments (if any). It encompasses the arguments associated with the operator, if applicable.
    \item From 0 to $m$ input token(s) (variable names or literals).
\end{enumerate}

Jaxpr is a strongly typed language but not the produced Python code. This is why the variable types are ignored at the tokenization step.

Some Jaxpr variable names may be forbidden by the Python language. A set of all forbidden variable names is stored (e.g. ``if'', ``in'', ``is'') to identify forbidden variable names and replace them with the uppercase version without the risk of colliding with the keywords.

\subsection{Line Translator} 

% Rules
JaxDecompiler's Line Translator contains a set of functions taking Jaxpr line represented as tokens and producing an equivalent Python string code. When a Jaxpr operator is missing, a clear Python Exception indicates the Jaxpr operator name to allow understanding and invite the community to implement it. Over 70 Jaxpr operators have been implemented, tested, and already addressing diversified applications.

We may enumerate some implemented Jaxpr operators in 3 categories:
\begin{itemize}
    \item Element-wise: `+', `*', `-', `/', `and', `or', `cos', `sin', `tan' ...
    \item Tensor manipulation: dot, transpose, convolution, sort ...
    \item High-order functions: condition, scan, parallel map, vectorized map ...
\end{itemize}

The Line Translator handles high-order instructions by applying recursive calls inside operator settings. In the Jaxpr language, high-order functions are represented using lambda expressions (unnamed function) stored in the operator arguments. To enhance code maintainability and reusability, a named function is produced and named with an incrementing index.

\subsection{Import Set}

Some Jaxpr functions can be translated into native Python language, while others require additional modules. Each time a line is encountered the Line Translator adds the import instruction's string into the import set. The Import Set keeps track of necessary import statements, ensuring efficiency and avoiding duplicate imports. We assume that the order of import has no importance.

Ultimately, after reading and translating all Jaxpr code, and the Import Set has tracked the necessary imports, the output of the Import Set and the Python code are concatenated. This constitutes the final Python code.

\section{Performance of the decompiled code}
\label{perf}

The provided section gives an overview of the performance evaluation of code decompiled by JaxDecompiler. It aims to assess whether the decompiled Python code can maintain performance levels comparable to the original Jaxpr code.
The performance is scrutinized across five distinct applications:
\begin{itemize}
    \item Training with Multi-Layer Perceptrons (MLP): Involves computing the gradient of the neural network for backpropagation during training. Various settings are explored, including different numbers of data points, units per layer, and layers.
    \item Inference with MLP: Similar to the training scenario, performance is assessed while predicting MLPs under varying settings.
    \item Sorting using MapReduce: Utilizes multi-core CPU for sorting 32 million random numbers and returning the three smallest elements to the user. Specifically employs the multi-core "pmap."
    \item Molecular simulation: Involves simulating molecules represented as a 3D point cloud, utilizing gradient descent to update their positions at each time step to reach a stable equilibrium state.
    \item All reduce in multi-node multi-core settings: Computes an array of nine elements containing the average of process identifiers (MPI rank) based on the average allReduce collective communication operation in a distributed setting. Evaluated on the University of Luxembourg HPC \cite{ulhpc:2014} named Aion  \footnote{\url{https://hpc-docs.uni.lu/systems/aion/}}.  Notably, JAX users involved in data-parallel code use `mpi4jax' \cite{mpi4jax:2021}, while JaxDecompiler translates this with `mpi4py' \cite{mpi4py:2021}.
\end{itemize}
For each application, the results are presented based on the average and standard deviation time (in seconds) across ten runs. The CPU used is an AMD EPYC with 128 cores (without hyper-threading) which is a common CPU in computing-intensive infrastructures. The performance is summarized in Table~\ref{tab:perf1} for unjitted code and Table~\ref{tab:perf2} after Just-In-Time (JIT) compilation.

\begin{table}
\centering
\caption{Performance comparison (seconds) of unjitted JAX function before and after decompilation}
\label{tab:perf1}
\begin{tabularx}{\linewidth}{llllcccc}
Application                & \multicolumn{3}{c}{Settings}                      & JAX function            & Decompiled         \\ \hline
\multirow{3}{*}{Training}  & \multicolumn{3}{l}{1K points 1K units 2 layers}   & $13.49\pm0.5$       & $\mathbf{8.67\pm0.07}$      \\
                           & \multicolumn{3}{l}{16 points 1K units 128 layers} & $10.43\pm0.15$      & $\mathbf{5.3\pm0.1}$        \\
                           & \multicolumn{3}{l}{16 points 8K units 2 layers}   & $\mathbf{8.85\pm0.06}$       & $11.52\pm0.07$     \\ \hline
\multirow{3}{*}{Inference} & \multicolumn{3}{l}{1K points 1K units 2 layers}   & $\mathbf{0.3685\pm0.001}$    & $0.4351\pm0.0187$  \\ 
                           & \multicolumn{3}{l}{16 points 1K units 128 layers} & $\mathbf{0.4725 \pm 0.0008}$ & $0.635 \pm0.0405$  \\
                           & \multicolumn{3}{l}{16 points 8K units, 2 layers}  & $\mathbf{0.4546\pm0.0002}$   & $0.4681\pm0.0002$  \\  \hline
\multirow{5}{*}{Sorting}   & \multicolumn{3}{l}{1 core}                        & $\mathbf{23.6494\pm0.2312}$  & $24.5791\pm2.1247$ \\ 
                           & \multicolumn{3}{l}{2 cores}                       & $11.778\pm1.4836$   & $\mathbf{11.5528\pm1.3912}$ \\
                           & \multicolumn{3}{l}{16 cores}                      & $\mathbf{1.0613\pm0.0381}$   & $1.0812\pm0.0115$  \\
                           & \multicolumn{3}{l}{128 cores}                     & $\mathbf{0.2312\pm0.0246}$   & $0.2784\pm0.0384$  \\  \hline
\multirow{2}{*}{Physics}   & \multicolumn{3}{l}{10,000 iter. 2 molecules}      & $48.29\pm0.06$      & $\mathbf{17.95\pm0.09}$     \\ 
                           & \multicolumn{3}{l}{1,000 iter. 20 molecules}      & $92.04\pm0.17$      & $\mathbf{33.99\pm0.04}$     \\  \hline
\multirow{4}{*}{AllReduce} & \multicolumn{3}{l}{1 node 128 MPI ranks}          & $0.0439\pm0.0025$   & $\mathbf{0.0021\pm0.0006}$  \\ 
                           & \multicolumn{3}{l}{4 nodes 512 MPI ranks}         & $0.0749\pm0.0099$   & $\mathbf{0.0199\pm0.0074}$  \\
                           & \multicolumn{3}{l}{16 nodes 2048 MPI ranks}       & $0.0863\pm0.0076$   & $\mathbf{0.0243\pm0.0084}$  \\
                           & \multicolumn{3}{l}{64 nodes 8192 MPI ranks}       & $0.0998\pm0.0108$   & $\mathbf{0.0285\pm0.0064}$ 
\end{tabularx}
\end{table}

% \usepackage{multirow}
\begin{table}
\centering
\caption{Performance comparison (seconds) of JIT JAX function before and after decompilation}
\label{tab:perf2}
\begin{tabularx}{\linewidth}{llllcccc}
Application                & \multicolumn{3}{c}{Settings}                      & JAX function            & Decompiled          \\ \hline
\multirow{3}{*}{Training}  & \multicolumn{3}{l}{1K points 1K units 2 layers}   & $\mathbf{6.3\pm0.35}$        & $6.48\pm0.23$       \\
                           & \multicolumn{3}{l}{16 points 1K units 128 layers} & $5.92\pm0.05$       & $\mathbf{4.71\pm0.09}$       \\
                           & \multicolumn{3}{l}{16 points 8K units 2 layers}   & $\mathbf{11.06\pm0.04}$      & $12.38\pm0.07$      \\ \hline
\multirow{3}{*}{Inference} & \multicolumn{3}{l}{1K points 1K units 2 layers}   & $\mathbf{0.1378\pm0.0002}$   & $0.1622\pm0.002$    \\
                           & \multicolumn{3}{l}{16 points 1K units 128 layers} & $\mathbf{0.4307\pm0.0023}$   & $0.5792\pm0.0016$   \\
                           & \multicolumn{3}{l}{16 points 8K units, 2 layers}  & $\mathbf{0.442\pm0.0001}$    & $0.6238\pm0.0002$   \\ \hline
\multirow{5}{*}{Sorting}   & \multicolumn{3}{l}{1 core}                        & $24.1114\pm1.1893$  & $\mathbf{23.7351\pm0.0911}$  \\
                           & \multicolumn{3}{l}{2 cores}                       & $\mathbf{11.1945\pm1.1323}$  & $12.589\pm4.1843$   \\
                           & \multicolumn{3}{l}{16 cores}                      & $\mathbf{1.5066\pm0.3904}$   & $1.5164\pm0.3665$   \\
                           & \multicolumn{3}{l}{128 cores}                     & $\mathbf{1.6336\pm0.3475}$   & $1.706\pm0.1575$    \\  \hline
\multirow{2}{*}{Physics}   & \multicolumn{3}{l}{10,000 iter. 2 molecules}      & $\mathbf{6.36\pm0.02}$       & $9.79\pm0.03$       \\ 
                           & \multicolumn{3}{l}{1,000 iter. 20 molecules}      & $\mathbf{11.9\pm0.02}$       & $18.63\pm0.13$      \\  \hline
\multirow{4}{*}{AllReduce} & \multicolumn{3}{l}{1 node 128 MPI ranks}          & $\mathbf{0.0006\pm0.0001}$ & $0.0021\pm0.0001$   \\ 
                           & \multicolumn{3}{l}{4 nodes 512 MPI ranks}         & $0.0191\pm0.0058$   & $\mathbf{0.0017\pm0.0001}$ \\
                           & \multicolumn{3}{l}{16 nodes 2048 MPI ranks}       & $0.0247\pm0.0049$   & $\mathbf{0.0177\pm0.0048}$ \\
                           & \multicolumn{3}{l}{64 nodes 8192 MPI ranks}       & $0.0334\pm0.0137$   & $\mathbf{0.0277\pm0.0043}$
\end{tabularx}
\end{table}

The showcased applications underscore the versatility of decompiling various types of applications. In summary, the performance evaluation of the decompiled code demonstrates reasonable performance when compared to the original Jaxpr code. In the AllReduce scenario, the superior performance of the decompiled code is attributed to the direct nature of mpi4py in calling MPI (Message Passing Interface) primitives, as opposed to mpi4jax, which relies on mpi4py before reaching the MPI library. This additional layer of abstraction contributes to the observed performance differences. Additionally, the resilience in retaining the benefits of parallel and JIT instructions post-decompilation enhances the adaptability of the decompiled code for diverse performance-critical applications.

For the sake of transparency and reproducibility, we provide URLs at the document's end, offering access to JaxDecompiler's main code, the benchmarks used, and comprehensive tests. These resources serve as references for researchers and practitioners seeking to replicate and delve deeper into our study.

\section{Conclusion}
\label{conclusion}

In the ever-evolving landscape of numerical frameworks and gradient-informed software development, JAX has emerged as a versatile and performant framework. JaxDecompiler plays a pivotal role in reverse engineering machine learning functions generated by JAX, addressing a critical gap and empowering researchers to gain deeper insights into the inner workings of these functions. By offering a clearer and more accessible Python representation of Jaxpr code, JaxDecompiler facilitates debugging and analysis, crucial for identifying and addressing issues or unexpected behaviors. Furthermore, the software provides users with the capability to manually optimize the generated Python code, enhancing performance and arithmetic stability.

Notably, JaxDecompiler's performance aligns with that of code originally written, showcasing its effectiveness. While decompilers are inherently dependent on source and target language versions, JaxDecompiler stands as an open-source project, welcoming community contributions and remaining adaptable in the dynamic landscape of gradient-based software development and research.

Codes are available on GitHub: \url{https://github.com/PierrickPochelu/JaxDecompiler/}

\section*{Acknowledgment}

The experiments presented in this paper were carried out using the HPC platform of the University of Luxembourg. Special thanks to Florian Felten for its review.

%TODO:  The experiments presented in this paper were carried out using the HPC facilities of the University of Luxembourg (Varrette et al., 2014) – see https://hpc.uni.lu

%Hence, it's understandable that JaxDecompiler does not present a panacea due to the fact it should continually follow JAX and Jaxpr evolution. However, we anticipate active engagement from the community in discussions aimed at its ongoing refinement and enhancement.

%Despite its undeniable value, it's crucial to acknowledge that JaxDecompiler, while functional, may require effort to improve the code. Furthermore, the effectiveness of JaxDecompiler hinges on its alignment with updates in JAX, JAXPR, and Python, necessitating maintenance and updates to ensure its continued relevance. Nevertheless, JaxDecompiler's role as an open-source project invites collaboration and contributions from the community. The open source ensures its adaptability and utility in an ever-changing landscape, where gradient-based software development is at the forefront of innovation and research.

%\bibliographystyle{IEEEtran} % IEEE
\bibliographystyle{splncs04} % Springer
\bibliography{bib_decomp.bib}

\end{document}